\documentclass[5p,12pt]{elsarticle} 

\usepackage{amssymb}
\usepackage{hyperref}
\usepackage{comment}
\usepackage[utf8x]{inputenc}
\usepackage{textcomp}

\newcommand\blfootnote[1]{%
  \begingroup
  \renewcommand\thefootnote{}\footnote{#1}%
  \addtocounter{footnote}{-1}%
  \endgroup
}

\journal{Nucl. Instr. Method A}

\begin{document}

\begin{frontmatter}



\title{A Flexible Data Acquisition System Architecture for the Nab Experiment}


\author[1,2]{D. G. Mathews}
\ead{mathewsdg@ornl.gov}

\author[2]{H. Acharya}
\author[2]{C. B. Crawford}
\author[2]{M. H. Gervais}
\author[2,3]{A. P. Jezghani}
\author[2]{M. McCrea}
\author[2]{A. Nelsen}

\author[4]{A. Atencio}
\author[5]{N. Birge}
\author[1]{L. J. Broussard}
\author[6]{J. H. Choi}
\author[1]{F. M. Gonzalez}
\author[7]{H. Li}
\author[8]{N. Macsai}
\author[8]{A. Mendelsohn}
\author[8]{R. R. Mammei}
\author[9]{G. V. Riley}
\author[5]{R. A. Whitehead}

\address[1]{Physics Division, Oak Ridge National Laboratory, 1 Bethel Valley Road, Oak Ridge, 37830, TN, USA}
\address[2]{Department of Physics and Astronomy, University of Kentucky, Lexington, 40506, KY, USA}
\address[3]{Georgia Institute of Technology, Atlanta, 30332, GA, USA}
\address[4]{Drexel University, 3141 Chestnut St, Philadelphia, 19104, PA, USA} 
\address[5]{University of Tennessee, Knoxville, 37996, TN, USA}
\address[6]{North Carolina State University, Raleigh, 27695, NC, USA}
\address[7]{University of Virginia, Charlottesville, VA, USA}
\address[8]{University of Manitoba, Winnipeg, MB R3T 2N2, Canada}
\address[9]{Los Alamos National Laboratory, Los Alamos, 87545, NM, USA}

\begin{abstract}
The Nab experiment will measure the electron-neutrino correlation and Fierz interference term in free neutron beta decay to test the Standard Model and probe Beyond the Standard Model Physics. Using National Instrument's PXIe-5171 Reconfigurable Oscilloscope module, we have developed a data acquisition system that is not only capable of meeting Nab's specifications, but flexible enough to be adapted \textit{in situ} as the experimental environment dictates. The L1 and L2 trigger logic can be reconfigured to optimize the system for coincidence event detection at runtime through configuration files and LabVIEW controls. This system is capable of identifying L1 triggers at at least $1$\,MHz, while reading out a peak signal rate of approximately 2\,GB/s. During commissioning, the system ran at a sustained readout rate of $400$\,MB/s of signal data originating from roughly $6$\,kHz L2 triggers, well within the peak performance of the system.
\end{abstract}


\end{frontmatter}
\section{Introduction}\label{sec:Intro}
The Standard Model of Physics is a detailed framework that describes the fundamental forces and the mediating particles that make up the universe.\blfootnote{Notice: This manuscript has been authored by UT-Battelle, LLC, under contract DE-AC05-00OR22725 with the US Department of Energy (DOE). The US government retains and the publisher, by accepting the article for publication, acknowledges that the US government retains a nonexclusive, paid-up, irrevocable, worldwide license to publish or reproduce the published form of this manuscript, or allow others to do so, for US government purposes. DOE will provide public access to these results of federally sponsored research in accordance with the DOE Public Access Plan (http://energy.gov/downloads/doe-public-access-plan).}Several recent measurements indicate a deviation from Standard Model predictions, indicating an incomplete or misunderstood interpretation of physics \cite{pdg-overview}. While experiments at the Large Hadron Collider continue to push the boundaries of our knowledge of the Standard Model, low-energy endeavors offer complementary and competitive insights into non-standard phenomena. Investigations of decays such as free neutron beta decay can provide a more thorough understanding of the electroweak interaction and serve as a sensitive probe for Beyond the Standard Model (BSM) Physics \cite{CIRIGLIANO2023137748, brodeur2023nuclear}. 

Combined with the neutron lifetime, $\tau_{n}$, angular correlations such as the electron asymmetry or the electron-neutrino correlation can be used to determine the Cabibbo Kobayashi Maskawa matrix element $V_{ud}$ without the theoretical uncertainties present in many-bodied interactions such as superallowed $0^{+}\rightarrow0^{+}$ decays. Alternatively, a 1/$E_{e}$ dependence present in the neutrino asymmetry or a distortion to the beta spectrum, known as Fierz interference, is indicative of new physics in the form of scalar or tensor couplings\cite{RevModPhys.83.1111}.

\subsection{The Nab Experiment}
The Nab experiment at the Spallation Neutron Source (SNS) at Oak Ridge National Laboratory is the latest in a series of precision measurements of angular correlations in free-neutron beta decay. The goal of the experiment is to measure the electron-neutrino correlation term to a relative precision of $1\times 10^{-3}$ by reconstructing the decay kinematics from electron energy and proton momentum. An additional measurement of the Fierz interference term to an absolute precision of $3\times 10^{-3}$ is intended as a probe for BSM Physics. \cite{Pocanic2009}

\begin{figure}[!b]
\centering
\includegraphics[width=3.5in]{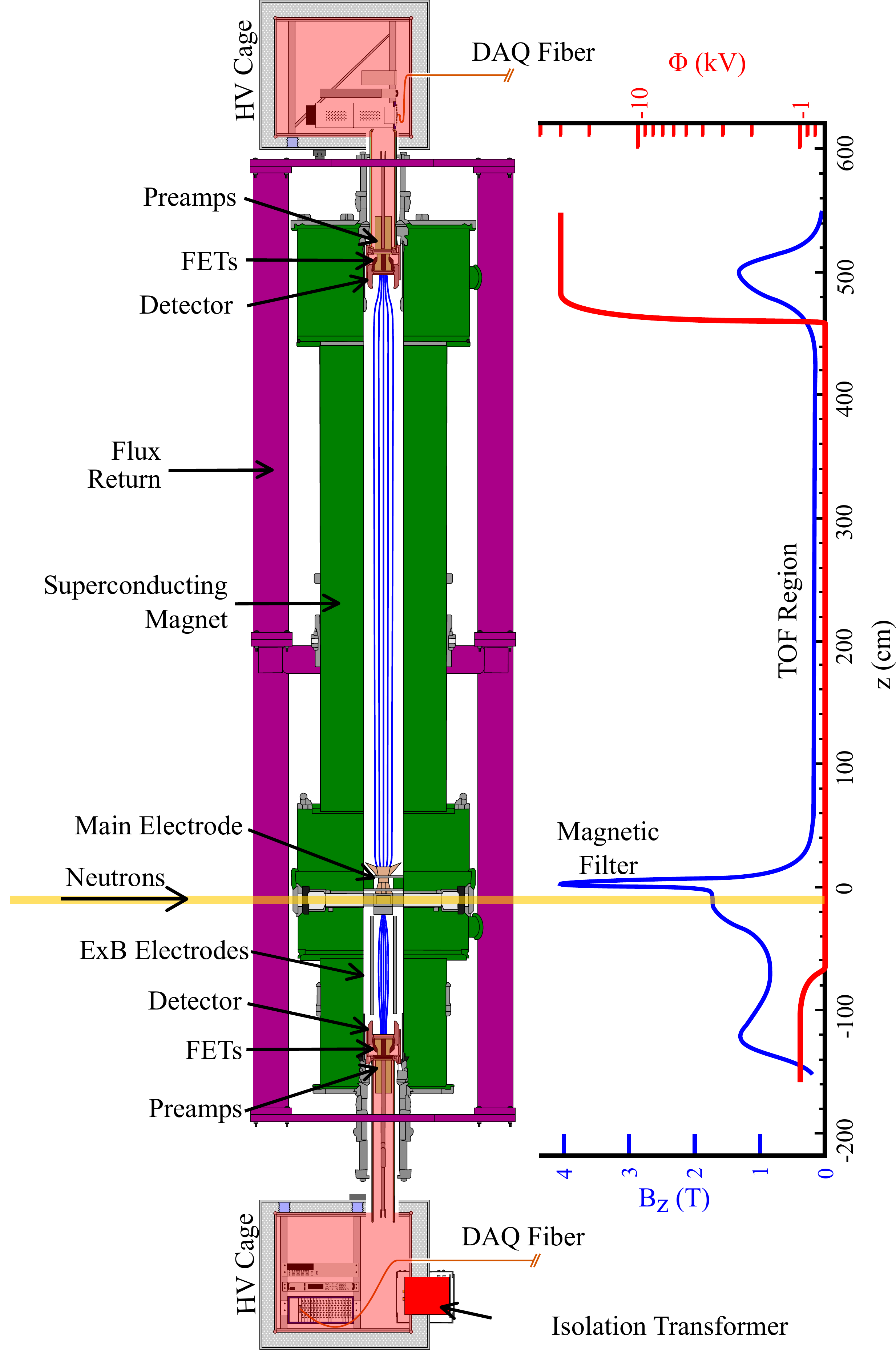}
\caption{The Nab spectrometer and data acquisition systems. Neutrons that decay while passing through the spectrometer are magnetically guided to detectors above or below the beamline. The electrons carry sufficient energy to deposit measurable signals, while the protons undergo electrostatic acceleration (requiring fiber-optic readout of the electronics) to gain detectable energies. Proton energies are thus determined via time-of-flight measurement.}
\label{fig:nab_model}
\end{figure}

Two 2.0 mm silicon wafers, divided into 127 hexagonal pixels, will measure, in coincidence, the electron and proton from unpolarized neutron decay. The electron energy can be determined directly from the charge deposited into the detector, while the proton energy must be determined via a time-of-flight measurement along a fixed-length spectrometer. Furthermore, the massive proton is unable to penetrate the 100\,nm dead-layer of the silicon with measurable energy, so a 30\,kV accelerating potential is applied in front of the detector to allow for successful detection \cite{PhysRevC.107.065503}. Figure \ref{fig:nab_model} depicts the experiment as installed at the SNS.

The data acquisition system (DAQ) for this experiment needs to support electron energy extraction with 3\,keV resolution with an integral non-linearity understood to the 0.01\% level. Additionally, any sytematic biases in timing extraction, such variations in clock synchronization, must be understood to better than $300$\,ps for proper phase space reconstruction for electron-proton coincidences, and the sample rate must be sufficient to capture multiple points on the $30-40$\,ns rising edge of detector signals for backscattering identification. Lastly, a low-threshold trigger must be able to efficiently detect protons, and flexible multipixel-logic must be implemented between the two detectors, across the 30 kV potential, for precise event reconstruction. Preliminary versions of this system were debugged and validated during the Calcium-45 and UCNB experiments at Los Alamos National Laboratory \cite{Ca45Reference, BROUSSARD201783, 10.1063/1.4826741, McGaugheyPL2009Motn}. 

\section{DAQ Hardware}
\begin{figure}[ht]
\centering
\includegraphics[width=3.5in]{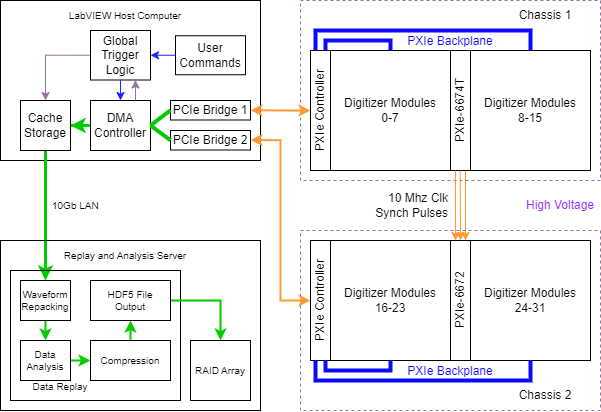}
\caption{The full system map for the Nab-DAQ. The two remote chassis generate L1 triggers and communicate these via fiber to the host computer where they are then processed to identify L2 triggers. From these L2 triggers, signals are requested from the two remote chassis and exported to the Cache Storage alongside L1 and L2 trigger information. A replay script then fetches this data from the cache, runs additional analysis, compresses, and then outputs the data to a RAID storage system.}
\label{fig:computer_map}
\end{figure}

The Nab-DAQ is intended to serve as a flexible system that can be easily modified for optimized performance in the event of unanticipated noise or a desire to modify the operating parameters of the experiment. With this in mind, the system was built using National Instruments hardware and was programmed using LabVIEW and LabVIEW FPGA. This system is comprised of two remote chassis that are responsible for the initial parsing of incoming data and identification of level 1 (L1) triggers. These L1 triggers are then sent to the host computer for parsing to identify different classes of events, or L2 triggers. Each L1 trigger, L2 event, and corresponding signal data is then output to a separate computer for replaying and post-processing. An overview of this system can be seen in Figure \ref{fig:computer_map}. 

\subsection{Digitizer Cards}
\begin{figure}[ht]
\centering
\includegraphics[width=3.5in]{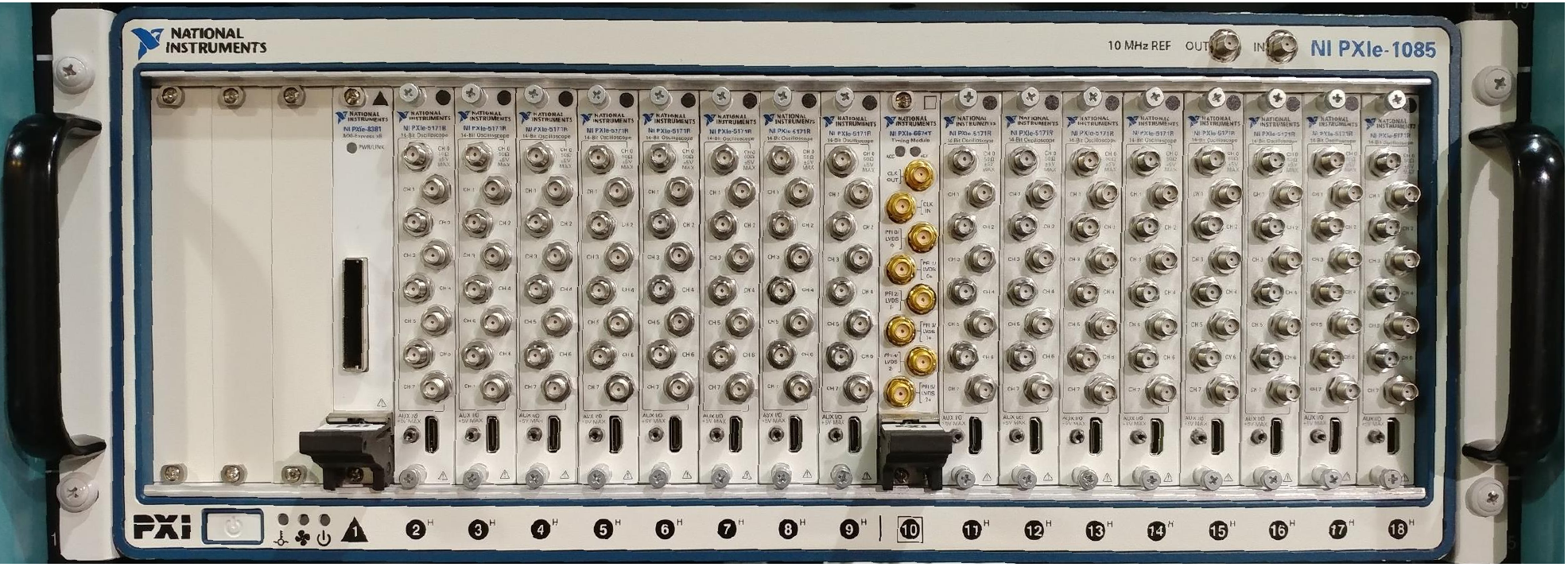}
\caption{One fully instrumented chassis for the DAQ. Slot 1 is the PXIe fiber controller, slots 2-9 and 11-18 are the digitizer cards, and slot 10 is the global OCXO reference.}
\label{fig:daq_chassis}
\end{figure}

The DAQ is designed around the PXIe-5171 Reconfigurable Oscilloscope cards \cite{PXIe5171}. Each card is comprised of an AD9250 ADC, offering 8 channel readout of 14-bit samples at 250MHz, backed by a Kintex-7 410T FPGA. The cards are run with the 120 MHz anti-aliasing filter and DC-coupling settings enabled and support full-scale voltage ranges from $0.2V_{pp}$ to $5V_{pp}$. Directly adjacent to the FPGA are two banks of DRAM, totaling 12 Gbit/card. Sixteen cards are installed in each of two 18-slot chassis, models PXIe 1085 and PXIe 1095. Each chassis is responsible for one detector; Figure \ref{fig:daq_chassis} shows one fully instrumented chassis. These chassis communicate with the host computer over MXIe Gen 2 fiber optical cables, a necessity due to the $30$\,kV potential difference between the computer and high-voltage cages depicted in Figure \ref{fig:nab_model}.

These FPGA modules are ideal for this application as they directly support the requirements of the experiment, in particular the sampling rate and non-linearity specifications. The 250MHz sampling rate of these cards allows for multiple data points to be collected along the rising edge of each signal, an essential feature for identifying particles that backscattered from the detector within a signal. Additionally, the AD9250 ADC has a reported maximum differential non-linearity (DNL) of $\pm0.75$\,lsb and maximum integral non-linearity (INL) of $\pm3.5$\,lsb with typical values of $\pm0.25$\,lsb and $\pm1.5$\,lsb respectively at $25^o$\,C \cite{AD9250}\footnote{Note that `lsb' stands for least significant bits.}. A typical integral non-linearity of $1.5$\,lsb for a 14-bit digitizer means that the output values are within $1.5/2^{14}\approx0.009\%$ of the true value. This is comparable to the $0.01\%$ level goal of the Nab experiment. As the maximum value per the manufacturer correlates to a $0.02\%$ non-linearity, a precision pulse generator with a rated linearity at the $0.0015\%$ level will be used to test each input channel on the FPGA modules in series with the detector electronics during the measurement cycle \cite{PB5Pulser} to ensure understanding of the non-linearity of the complete system to the $0.01\%$ level.

\subsection{Timing and Synchronization}
\begin{figure}
    \centering
    \includegraphics[width=\linewidth]{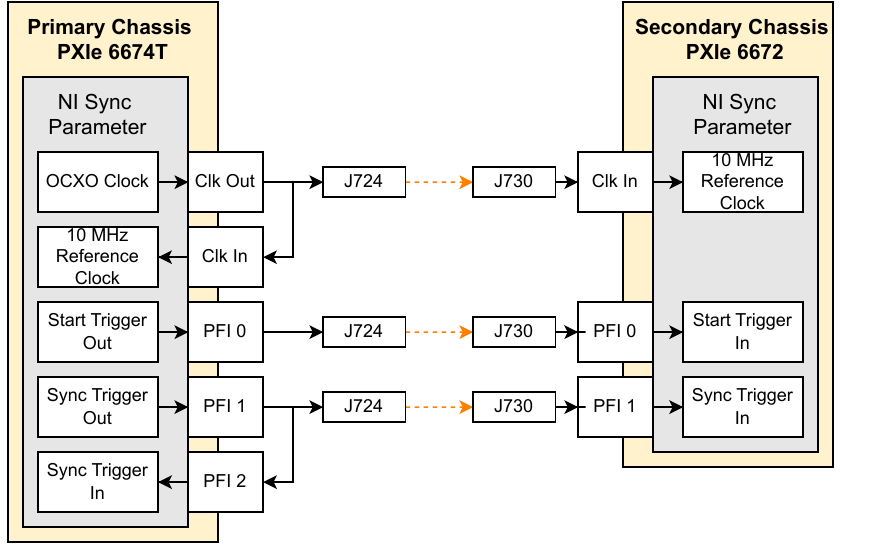}
    \caption{The hardware and software connections used for DAQ timing and synchronization. Solid black lines indicate electrical(software) connections, and dotted orange lines indicate fiber-optic connections. The Highland Technology J724+J730 combination is rated for $<$12ps RMS jitter\cite{HighlandTechnologyJ724}.}
    \label{fig:SynchronizationConfiguration}
\end{figure}

Each chassis is configured with a dedicated timing and synchronization module. The primary chassis is outfitted with a PXIe-6674T module that produces a 10MHz reference clock from its internal oven-controlled crystal oscillator. This clock is distributed across the primary chassis' backplane and to the secondary chassis using Highland Technology converters to alternate between electrical and fiber-optic signals to cross the high voltage potential. The PXIe-6672 module in the secondary chassis ingests the clock produced in the primary chassis and distributes it to the secondary chassis backplane. In addition to the clock, two additional connections are made between the PXIe-6674T and PXIe-6672 modules to support the alignment of each FPGA clock between the chassis\cite{SYNCWP}. This configuration is shown in Figure \ref{fig:SynchronizationConfiguration}. In tandem with the PXIe-5171 oscilloscope modules, the rated synchronization capability of this system is $300$\,ps, before manual adjustment, and $\leq10$\,ps with manual adjustment \cite{PXIe5171}, well within the Nab experimental precision goal of $300$\,ps. 

\subsection{Host Computer}
The host computer is responsible for management of the DAQ hardware and processing incoming data from the FPGA. Each chassis is connected to a PCIe bridge through the x8 PCIe Gen 2 interface giving a total system throughput of up to 4 GB/s/chassis. The cache storage on this system is an Intel Optane 905p 960GB drive chosen specifically for its high I/O performance, 2600/2200 MB/s read/write respectively, and high endurance rating, 17.52 PB written\cite{IntelOptane}. All data is transferred from this computer over 10Gb Ethernet to the replay server.

\subsection{Replay Server}
The replay server is configured with an AMD Threadripper Pro 5965WX 24-Core CPU with 128 GB of DDR4 memory and an Nvidia RTX A4000 GPU to support any post-processing and data replay that is desired. For data storage, an external storage array is configured with 44 Seagate EXOS 20TB drives, model ST20000NM003D. These drives are split into a pair of RAID 6 configurations, each with 400 TB of configured storage. These RAIDs are network mounted so additional local analysis machines can access the data. This allows shift takers to perform analysis on incoming data without over-utilizing the replay server or Host DAQ computer.

\subsection{Data Distribution and Long-Term Archival}
Data distribution is performed with the ORNL Physics Division cluster. All data saved to the replay server, after compression, is also transferred to a $300$ TB storage array configured on the Physics Division cluster. This is intended to store data from recent measurement cycles so that it can be broadcast to collaborating institutions, not for permanent backup. Users can either perform analysis directly on that cluster, or use resources such as Globus\cite{globus1,globus2} to transfer the data to external resources.

\section{Firmware and Software}

\subsection{FPGA Firmware and L1 Logic}
\begin{figure}[ht]
\centering
\includegraphics[width=3.5in]{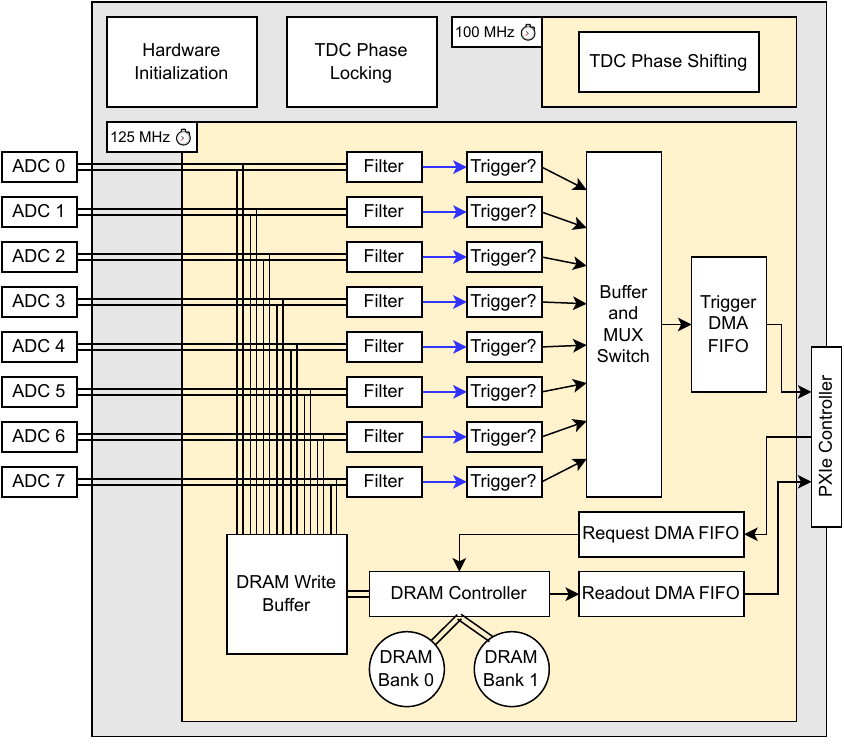}
\caption{A representation of the FPGA logic and data flow through the chip. As data is read in, the stream is written to a continuous ring buffer and also piped through a filtering and trigger logic system. Any identified triggers are sent over the Trigger DMA FIFO to the user-defined logic system where they are parsed. Signal requests generated by this logic are sent back to the FPGA over the Request DMA FIFO and the requested data is read out by the DRAM Controller logic from the pair of DRAM memory banks and sent back to the computer over the Readout DMA FIFO.}
\label{fig:fpga_logic}
\end{figure}

The firmware onboard the FPGA can be divided into three main parts: hardware initialization, data acquisition and processing, and signal output. The hardware initialization loop establishes the FPGA resources for routine running, such as clocking and communication with the PXIe backplane, as well as the system synchronization process \cite{SYNCWP}. A high-level overview of the FPGA firmware can be seen in Figure \ref{fig:fpga_logic}

The ADC operates at 250MHz while the data acquisition loop processes at 125MHz, so 2 samples are read from each channel every clock cycle. These data streams are then forked and sent to two different operations. One stream is structured into a binary sequence and written to a pair of DRAM memory banks. Each bank is configured with a 384 bit bus, allowing for 768 bits effectively per address. This system alternates between read and write modes to the DRAM banks to prevent race conditions occurring in the readout logic. Incoming data is held in a First-In First-Out (FIFO) buffer, or DRAM Write Buffer, until the DRAM controller enters write mode. The data is organized with each address containing 48 sequential samples from the same input channel. The DRAM controller is configured such that it alternates between reading and writing modes at fixed intervals, for example a 50 clock cycles in write mode followed by 10 in reading mode. The DRAM Write Buffer is responsible for buffering the incoming data while the system is in read mode, and the duration of the read/write modes of the arbitration must be carefully tuned to ensure that the FIFO never overflows. A configuration of 9 ticks in write mode, 5 in read mode, has been found to be optimal for this use case.

The second copy of the incoming sample stream is filtered by a modified trapezoidal filter \cite{JORDANOV1994337} and the filtered output is then analyzed to identify any signal that match the L1 trigger criteria. For each identified trigger, the system outputs the extracted energy and arrival time of physics events, the pixel the event was seen on, and a counter that keeps track of how many triggers have been identified from each channel \cite{jezghani_2019} \cite{david_2022} since the system was initialized. This counter helps ensure that offline analysis is aware of any loss of data that occurs between identification and output from the host logic system by flagging any skips in that counter. These triggers are then transferred to the host CPU via a direct memory access (DMA) FIFO, where they are then processed by the L2 trigger logic.

A major feature of this system is the ability to reconfigure the L1 trigger logic and signal processing to match the incoming data. The modified trapezoidal filter routine is configurable such that all channels have independently tuned decay rates, flat top lengths, rising edge lengths, dead times, and trigger thresholds. Trigger thresholds and dead times can be adjusted at run time, but the decay rate, flat top, and rising edge lengths must be set at initialization due to the variable shift registers that require pre-allocation of memory resources. 

The entirety of the trigger logic and signal processing operations take place within two function calls on the FPGA, one for each process. This design allows for the system to be reconfigured to utilize other filtering methodologies with a minimally invasive reprogramming process. Any adjustments to these functions will require re-synthesis of the FPGA firmware, but the remaining architecture such as the signal readout operations do not need to be adjusted to accommodate these changes.

\subsection{LabVIEW Software and L2 Logic}
The L1 triggers produced by the FPGA processing are then passed via DMA FIFOs to the host CPU, where they are aggregated and time-ordered. The L2 logic is implemented in LabVIEW software, using configuration files to dictate logic criteria. Incoming L1 triggers from all channels are aggregated and time sorted before being parsed to identify electron-proton coincidences based on timing, energy, and input channel information recorded in each trigger. In addition to coincident proton-electron events, additional L2 triggers are created in this process such as periodic sampling of both detector arrays in their entirety for analysis of baseline system noise, bias pulsing of the detector for gain monitoring, temperature sensor readout on the FPGA and ADC boards for systematic studies, and non-coincidence triggers for studying backgrounds are also determined in this process. Each L2 trigger includes information about the L1 trigger(s) associated with it, when applicable, as well as a run-specific unique identifier. All L1 and L2 trigger information is recorded in corresponding datasets within the output HDF5 file for post-processing. Any detector data associated with an L2 trigger is requested from the FPGAs by sending a request containing the desired channel, timestamp, and request length over DMA FIFO to the corresponding FPGA.

These requests are additionally sent over LabVIEW queue communication protocol to a multi-threaded signal retrieval loop on the host system. This loop pulls signal data from each FPGA via DMA FIFO, checks the signal header against the copy of the requests that was sent via LabVIEW queue to verify the expected signal was returned, and saves these headers and signals to a flat binary file structure. Each parallel instance of this loop, one for each FPGA, has its own dedicated output file that will need to be parsed by the replay script in order to combine all the data together. This parallel file writing is part of the motivation for the Intel Optane 960p cache SSD as it has excellent random read/write performance compared to traditional drives and is capable of keeping up with the high data rate of this system. Previous iterations of this system had each FPGA writing to different datasets in the same HDF5 file that the L1 and L2 triggers are saved to, but in testing that proved to have lower throughput than this parallel methodology. In addition to writing the files in parallel, multiple signals are retrieved by each DMA FIFO request to maximize throughput from the FPGAs as each transfer has a non-zero amount of overhead associated with it. 

\subsection{Analysis and Storage Server}
The data output from the LabVIEW host computer is saved initially to the cache drive as a combination of an HDF5 file containing configuration parameters and other system meta-data while the L1, L2, and signal data are all saved to their own dedicated binary files for each data type \cite{hdf5}. This splitting of file types was determined to be the optimal configuration to obtain peak throughput for the signal readout while saving the configuration data in a flexible format. A script copies this data to a temporary buffer on the RAID array where the replay process then parses the files. Once each file is loaded into memory, check-sums are calculated for each signal and inserted into their headers so offline analysis processes can verify data integrity. This is performed specifically at this stage as immediately after these are calculated, the headers and signals are split and saved to distinct storage locations within the HDF5 file based on the type of L2 trigger they are associated with. These checksums ensure that the analyzer can verify which header matches to each signal in case of an error occurring in this process, or during any future modifications to the files such as data compression. This split is done to improve offline analysis performance as many analysis operations only access header information and don't need to access the signal data and vice-versa. The signal data is compressed using the DeltaRice algorithm and saved to the HDF5 file along with any analysis results \cite{Mathews2024}. These data files are saved to the local RAID arrays for short term storage and local analysis efforts. 

To enable more rapid feedback on the system, additional analysis can be enabled through the nabPy library: \url{https://pypi.org/project/nabPy/}. This additional analysis creates secondary datafiles that contain the results of the analysis along with a full copy of the original data file, excluding the raw signal data. These smaller analysis result files are significantly faster to distribute to the analyzers. This pre-processing leverages the Nvidia RTX A4000 in the system through the CuPy library to accelerate the analysis. The full replay output, including both the original data file and the pre-processing output file, is sent to the ORNL Physics Division storage server for data distribution.

\subsection{Auto-Calibration DAQ}
In addition to the standard operational DAQ mode, the Nab-DAQ also offers an auto-calibration mode to assist the user with determining the optimal L1 trigger threshold on a per-channel basis. This system automatically iterates through all of the configured channels and determines the trigger rate as a function of the threshold parameter. In the standard running configuration, the various L1 triggers are returned to the host computer and parsed through the L2 logic. In this configuration, only the number of triggers returned is recorded per measurement interval with the triggers themselves being discarded and the rest of the L2 logic bypassed. With the trigger rates as a function of threshold determined, the system fits each distribution to a Gaussian function and places the trigger threshold at $3\sigma$ from the mean by default, though this can be adjusted by the user. The resulting trigger rate is saved to an output HDF5 formatted file for the user to parse manually if desired. 

\subsection{NI-Scope Soft Front Panel}
NI-Scope Soft Front Panel is a software suite provided for the PXIe-5171 oscilloscope modules that allows for the user to have oscilloscope-like control over each channel. This software does not use the same L1 or L2 trigger logic as the Nab-DAQ, but it does allow for the user to observe incoming detector signals in a similar fashion to a standard benchtop oscilloscope. Included in this software is the ability to only view certain channels, change trigger types to modes such as rising or falling edge detection, adjust trigger thresholds, save images, and more. This software suite allows for the Nab-DAQ system to effectively serve as a 256 channel benchtop oscilloscope, which is particularly useful for tasks such as debugging connectivity to detector channels during installation of the detector system. 

\section{System Performance}
The performance of the Nab-DAQ system is examined in this section. The system throughput, in terms of supported trigger rate and readout rate, was examined relative to the theoretical maximum throughput determined by the various interfaces used, such as PCIe and SSD read/write performance. Additionally, the timing synchronization performance of the DAQ is examined to determine how the system contributes to the overall Nab uncertainty budget. The linearity of the system is not explored within this document, beyond the manufacturer specifications, as the linearity of both the DAQ and detector electronics will need to be examined in tandem for systematic studies.

\subsection{Data Throughput}
With the first versions of the firmware, each FPGA was capable of continuous data transfer rates of approximately $130$ MB/s \cite{jezghani_2019} when outputting four samples packed into each output container. The updates presented here leverage a newly restructured DRAM buffer and read/write arbitration system. In the current configuration, the maximum stable readout achieved was $260$\, MB/s per FPGA, fully saturating the available $4$ GB/s per chassis upper limit set by the PCIe Gen 2.0 interface between each chassis and the host computer.

The output data is initially saved to the cache SSD which has a peak write performance of around $2200$ MB/s. This presently sets the upper limit on the peak instantaneous data rate. The data is moved from this system over a $10$ Gb/s Ethernet communication between the LabVIEW host system and Analysis and Storage server, which restricts the maximum time-averaged data rate to $1.25$ GB/s. Any sustained rates over this limit may eventually fill the cache SSD and cause downtime while the cache flushes to the RAID storage server. This has been observed when performing benchmarks of the system, but not in standard measurement configurations where the data rate has ranged between $100$ and $500$ MB/s, well within these hardware limitations.

The system is capable of identifying over $1$\,MHz of L1 triggers across all $32$ FPGA boards and exporting this information to the cache SSD in real-time, so long as the L2 logic and signal readout is disabled. With L2 logic enabled, the system throughput reduces to around $35$ kHz of L1 triggers due to the additional overhead of identifying the various L2 triggers. This performance can be improved through more efficient logic and work is underway to address these deficiencies. The final stage of the host computer process, the signal readout, is limited in its throughput by $2$ main obstacles: the $260$ MB/s per FPGA transfer rate, and the host computers processing capabilities such as memory speed and CPU clock rate. Many adjustments have been made to mitigate these bottlenecks. In practice, the $260$ MB/s per FPGA transfer rate is not the primary limitation and software inefficiencies are far more relevant. Converting the signal request and readout operations to occur in parallel for loops with individual dedicated output files for each FPGA allow the system to more efficiently use the CPU and memory resources. In addition, transitioning the software from $32$-bit to $64$-bit allowed for larger memory buffers which increased both the system stability and peak throughput. With these changes in place, the signal readout reached a peak measured throughput of $\approx 2100$ MB/s when recording signal traces from all channels bypassing the L1 and L2 logic, just below the full write performance of the cache SSD. With the L1 and L2 logic enabled and in the expected experimental configuration, the LabVIEW system has been observed to run stably at approximately $300$\,MB/s with over $15$\,kHz of incoming L1 triggers.

Depending on the analysis enabled in the replay script, the performance of that process varies significantly. The signal reading from cache SSD, checksum calculation, and compression have each been demonstrated to perform in excess of $1000$ MB/s. The writing of data to the RAID array peaks at $\approx1000$ MB/s for small transfers, but falls to a continuous throughput of $\approx 800$ MB/s per RAID array. By alternating the target destination of the replay script, it is possible to write to them both simultaneously for a cumulative rate of $1600$ MB/s, though observed sustained data rates have not made this necessary.

The total throughput of the system is limited by the slowest component. In this case, the 10Gb networking between the cache SSD and the RAID array is the slowest hardware component. This could be avoided by moving the data compression process to the same system as the LabVIEW DAQ software. This was tested and in practice resulted in a more unstable system due to over-utilization of the available CPU and memory resources. With the expected experimental data rate of between $100$ and $400$ MB/s the $10$ Gb/s network bottleneck is well beyond the experimental requirements and the cache drive allows for a buffer in case of occasional peaks beyond $10$ Gb/s. 

\subsection{Clock Synchronization}
\begin{figure}
    \centering
    \includegraphics[width=\linewidth]{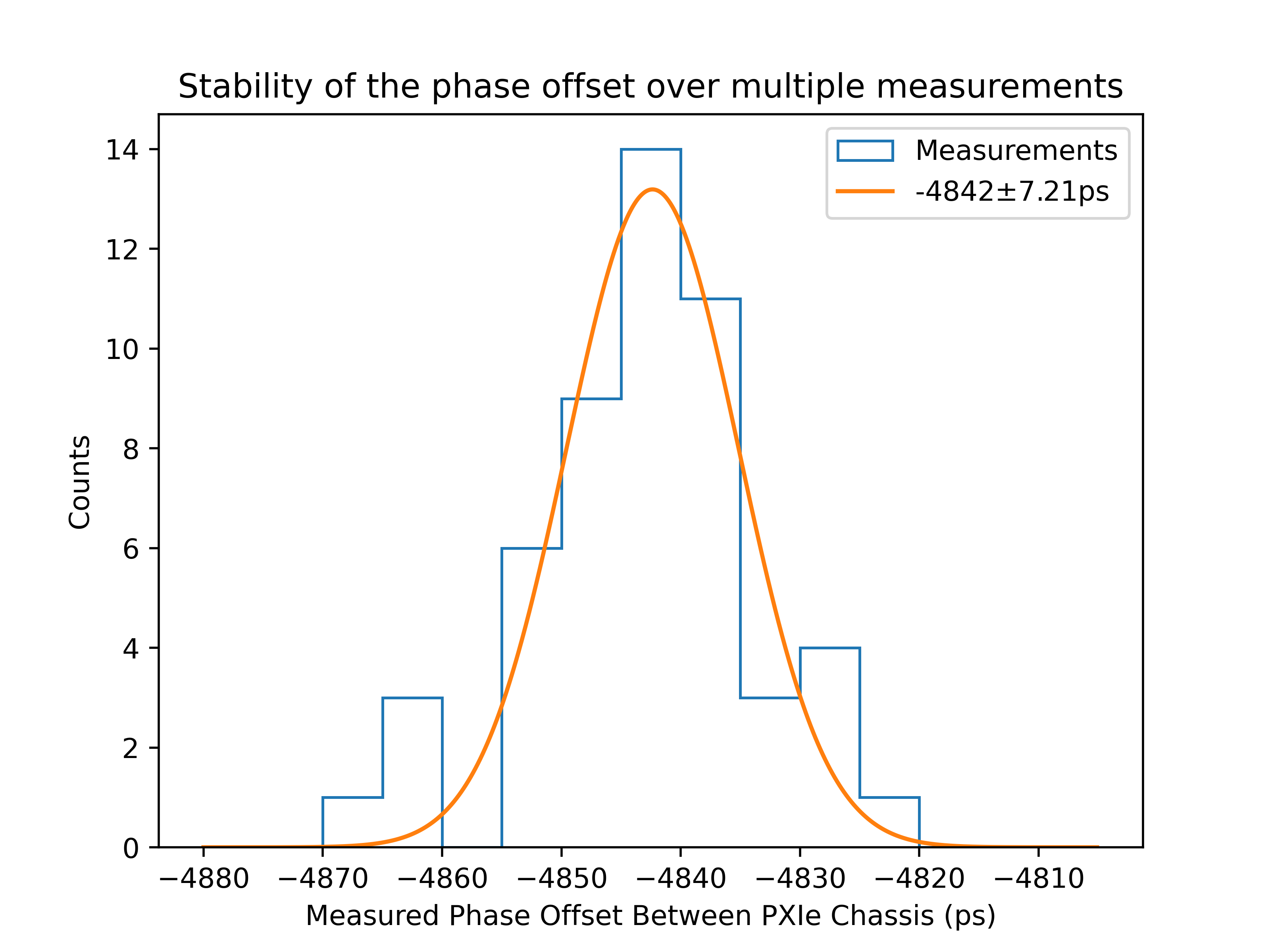}
    \caption{Demonstration of the stability of the synchronization over 50 repeated measurements of the system. The same configuration was used for each test to ensure no additional sources of uncertainty such as variations in the cable length. This test was performed with two input channels on FPGAs within different chassis with the synchronization signals being sent over fiber-optic connections as will be used in the full experiment. These measurements were fit to a gaussian function with the mean and standard deviation reported in the figure.}
    \label{fig:SynchronizationTesting}
\end{figure}

The alignment of the clock synchronization was studied by injecting an oscillating reference signal into pairs of channels across various FPGAs within the DAQ system. Whenever the DAQ software is started, any offset between the channels within the system must be consistent to at least $300$\,ps, and held stable to that level for the duration of the measurement period. The synchronization of the FPGA clocks, both on the same chassis and across the pair of chassis, is performed through the NI Tclk API \cite{SYNCWP}. Testing of this system was done by splitting a sinusoidal signal across multiple channels of the DAQ and measuring the deviation in the phase of these signals. Each signal was acquired by querying the timestamp on the first FPGA in the system, then requesting a signal from that timestamp in the ring buffer from every FPGA channel to ensure that each signal was requested from the same timestamp according to the DAQ system. These signals were then fit to a sine wave and the phase of each signal was determined. The deviation between the phases measured on different channels and its uncertainty determines the clock offset and uncertainty between FPGA channels. Figure \ref{fig:SynchronizationTesting} shows the results of one such test which indicates a stability at the $<10$ps level across two FPGA chassis, well below the Nab precision requirements and in agreement with the quoted performance of the NI Tclk API. 

The extraction of the start time from each signal can additionally contribute to timing jitter/uncertainty if there are biases present in the extraction method. This data acquisition system allows for the user to archive signals associated with L1 triggers so that they can be analyzed offline to correct for any such biases. This offline correction assumes a negligible bias originating in the trigger logic on the shape of the signals recorded which will need to be explored in more detail during the operation of the experiment. 

\section{Conclusion}
We have developed a novel DAQ architecture for the Nab experiment that is both highly adaptable to match the experimental configuration and also capable of exceeding anticipated experimental data rates. L1 trigger parameters, such as trigger threshold and shaping filter parameters, and the L2 trigger logic can both be adjusted at initialization/run-time allowing the user to adapt the behavior of the system to account for variations in the experimental conditions. This system is capable of exporting data to compressed HDF5 files in a local RAID storage server at up to $\approx 800$ MB/s. All L1 triggers, L2 logic results, and signal data are saved along with the configuration parameters and other relevant metadata to ensure that offline analysis has access to all relevant information regarding the configuration of the system. Additional analysis is made available at the replay stage through the nabPy library, including signal fitting routines and checksums to verify data integrity. 






\bibliographystyle{elsarticle-num}
\bibliography{main.bib}

\end{document}